\documentclass[onecolumn]{emulateapj}
\shorttitle{On the TFR }
\shortauthors{C. E. Navia}

\usepackage{amsmath}

\begin{document}

%% LaTeX will automatically break titles if they run longer than
%% one line. However, you may use \\ to force a line break if
%% you desire.

\title
%On the steep velocity-mass dependence at the faint end of the Tully Fisher relation
{
Dwarf galaxies at the steep and faint end of the Tully Fisher relation
}

%% Use \author, \affil, and the \and command to format
%% author and affiliation information.
%% Note that \email has replaced the old \authoremail command
%% from AASTeX v4.0. You can use \email to mark an email address
%% anywhere in the paper, not just in the front matter.
%% As in the title, use \\ to force line breaks.

\author{Carlos. E. Navia }
\affil{Instituto de F\'{i}sica, Universidade Federal Fluminense, 24210-346, Niter\'{o}i, RJ, Brazil}

%% Notice that each of these authors has alternate affiliations, which
%% are identified by the \altaffilmark after each name.  Specify alternate
%% affiliation information with \altaffiltext, with one command per each
%% affiliation.

\altaffiltext{1}{E-mail address:navia@if.uff.br}
%\altaffiltext{2}{Society of Fellows, Harvard University.}
%\altaffiltext{3}{present address: Center for Astrophysics,
%    60 Garden Street, Cambridge, MA 02138}
%\altaffiltext{4}{Visiting Programmer, Space Telescope Science Institute}
%\altaffiltext{5}{Patron, Alonso's Bar and Grill}

%% Mark off your abstract in the ``abstract'' environment. In the manuscript
%% style, abstract will output a Received/Accepted line after the
%% title and affiliation information. No date will appear since the author
%% does not have this information. The dates will be filled in by the
%% editorial office after submission.

\begin{abstract}
Nearby isolated galaxies ($z\sim 0$), are considered immerse within a thermal bath at 2.73 K. However the dwarf galaxies orbiting these galaxies are also subject to additional radiation from their hosts, so they are within a thermal bath slightly warmer.  We claim that this thermal effect can explain several properties of the dwarf galaxies, such as their rotation curves, their radial acceleration relations (RAR) and why the velocity-mass dependence at the faint end of the Tully Fisher Relation (TFR) is steeper for these galaxies. 
In the Debye Gravitational Theory (DGT), the galaxies properties, such as the rotation curves, the RARs
and the TFRs are isothermal curves; they depend explicitly only of temperature, of the thermal bath in which are immersed.
We show that the steep and faint end of the TFR is composed of a mixture of dwarf galaxies immersed in thermal baths with temperatures from 2.73 K to 3.80 K. A comparison among DGT's predictions for the dwarf galaxies relations with data obtained from the literature show a satisfactory agreement between them.
\end{abstract}

%% Keywords should appear after the \end{abstract} command. The uncommented
%% example has been keyed in ApJ style. See the instructions to authors
%% for the journal to which you are submitting your paper to determine
%% what keyword punctuation is appropriate.

%% Authors who wish to have the most important objects in their paper
%% linked in the electronic edition to a data center may do so in the
%% subject header.  Objects should be in the appropriate "individual"
%% headers (e.g. quasars: individual, stars: individual, etc.) with the
%% additional provision that the total number of headers, including each
%% individual object, not exceed six.  The \objectname{} macro, and its
%% alias \object{}, is used to mark each object.  The macro takes the object
%% name as its primary argument.  This name will appear in the paper
%% and serve as the link's anchor in the electronic edition if the name
%% is recognized by the data centers.  The macro also takes an optional
%% argument in parentheses in cases where the data center identification
%% differs from what is to be printed in the paper.


%% From the front matter, we move on to the body of the paper.
%% In the first two sections, notice the use of the natbib \citep
%% and \citet commands to identify citations.  The citations are
%% tied to the reference list via symbolic KEYs. The KEY corresponds
%% to the KEY in the \bibitem in the reference list below. We have
%% chosen the first three characters of the first author's name plus
%% the last two numeral of the year of publication as our KEY for
%% each reference.

\section{Introduction}
\label{sec:intro}

Dwarf galaxies are the most abundant type of galaxy in the universe. These faint galaxies have luminosities as low as $L \sim 1 \times10^{-3} L_{sun}$, and masses around two orders magnitude smaller than the spiral galaxies, like Andromeda. The classification of dwarf galaxies is in two types, the isolated dwarfs, but the more abundant, at least in the local field, are the satellite dwarfs, they orbit around a more massive companion, also known as host galaxy.
Isolate dwarf galaxies are typically too faint to be observed at high redshifts. 
According to \cite{prad02} the Local Group dwarf galaxies show strong correlations between mass-to-light
(M/L) ratio, surface brightness, and
metallicity, consequently, in the Local Group, we can obtain information on the chemical elements in the early star formation in these galaxies. 

 The classification of dwarf galaxies can also be according to their geometric form, in ellipticals, spheroidal and irregular. Elliptic dwarf galaxies are similar to the normal elliptical galaxies. However, they have lower metallicities and different light distributions. Spheroidal dwarf galaxies are fainter than the elliptic, and for this reason, they are observed only within the Local Group of galaxies. Finally, the dwarf irregular galaxies appear to have the same global properties seen in normal irregular galaxies. They are a cradle of stars because they have vast amounts of gas and dust.

Within the current most popular cosmological model:
the Lambda Cold Dark Matter ($\Lambda$CDM), the framework developed to describe massive spiral galaxies, has several inconsistencies when applies to dwarf galaxies from Milky Way and M31 (Andromeda) \citep{krou10} and recently these inconsistencies also were observed in the satellite galaxies around Centaurus A \citep{mull18}. $\Lambda$CDM  predict a population significantly higher
than the number of dwarfs orbiting these giant spiral galaxies. Also, the distribution of satellite galaxies is in a planar arrangement, and the members of the plane are circling in a coherent direction, this behavior contrasts with the $\Lambda$CDM predictions, where these satellites should move randomly around their host. 

The kinematic and dynamic of the rotation of dwarf galaxies, also constrains gravitational models, several of them based on $\Lambda$CDM model \citep{mada08,papo17,ferr12} need additional feedbacks to describe the dark matter merger in gas and to obtain an adequate scenario, to described the dwarf galaxies.
This scenario is also known as the small-scale problems, for instance, the dwarf spheroidal (dSph) galaxies from the Milky Way require small-scale calibrations to taken into account the observations.

On the other hand, modified gravity models, such as MOND \citep{fama12} gives clear and precise predictions on the dynamics of nearby galaxies, for instance, provides an accurate prediction on the baryonic Tully Fisher Relation (bTRF) \citep{mcga11},  a scaling-law relation between the baryonic-mass and velocity observed in nearby spiral galaxies. MOND also need of a feedback to describe the dwarf galaxies, the External Field Effect (EFE) mechanism \citep{brad00,mcga10b,mcga13}. This means that the gravitational dynamics of a dwarf galaxy is influenced by the external gravitational field in the which is embedded, in this case, its host galaxy. However, some predictions of this mechanism are hard to be observed, especially in the local dwarf spheroidals (dSph) galaxy data.

It is well known that the slope in TFR become increasingly steep at the faint end. To values of $v_{circ}$ around $\sim$100 km s$^{-1}$ there is a ``knee'' on TFR. 
\citep{mcga00,amor09,mcga10}. 
This steep faint end in the TFR is allocated by the dwarf galaxies and have several explanations, such as supernova feedback \citep{van00} and within the standard cosmological model $\Lambda$CDM it is claimed that this behavior arises from the mechanism called as the halo abundance matching (HAM), and dynamical corrections, such as the adiabatic halo contraction \citep{truj11} and small-scale calibrations \citep{sawa16}.  Also, this steep faint end, can be explained within the MOND paradigm including the EFE mechanism, and it would be responsible to put the dwarf galaxies out of equilibrium \citep{brad00,mcga10b}, breaking
the TFR scaling-law.

In this paper, we present another alternative to describe the dynamic of dwarf galaxies, inspired in DGT \cite{navi17,navi18}, a thermodynamic gravitational theory. DGT is not a modified gravity theory, it is built from first principles. DGT is an extension for low temperatures of the  Entropic Gravity Theory (EGT) \citep{verl11}. In this sense, EGT plays the role of Dulong-Petit law, for the specific heat of solids at high temperatures ($T>>T_D$) in the Debye theory of solids. In other words, DGT at high temperatures coincides with the Newtonian gravitational theory as well as through the Unruh relation is possible to show that is consistent with the General Theory of Relativity \citep{verl11}. However, at low temperatures ($T<T_D$), DGT coincides with the deep-MOND regime.
Here, $T_D=6.35$  K is the Debye temperature and constitute the free parameter of DGT; and it plays the role of Milgrom's acceleration scale ($a_0$) of MOND. DGT allows obtaining a bond between them.

In this framework the shape of rotation of a galaxy depends only on the thermal bath in which is immersed, this is also valid for dwarf galaxies. We show that DGT allows obtaining promising results describing the dwarf galaxies, such as their rotation curves, with only two free parameters, the observed galaxy mass, and a fit value for the thermal bath temperature. In this sense, in DGT the rotation curve is an ``isothermal curve''. Also, we show that both, RAR and the TFR predicted by DGT are also isothermal curves, both depending explicitly only of the temperature.

The organization of the paper is as follow. Section~\ref{background} 
is dedicated to a brief description of some main ingredients of DGT. The DGT predictions for rotation curves of dwarf galaxies are present in section~\ref{rotation}.
In section~\ref{radial} and in section~\ref{TFR} 
we present the DGT predictions for the RAR and the bTFR of dwarf galaxies, respectively.
Finally, in section~\ref{conclusions} we present our conclusions.

\section{Some main ingredients of DGT}
\label{background}

DGT follow the concept of  ``induced gravity'', initially proposed by Sakharov  (Visser, 2002). According to this concept, gravity is not ``fundamental'' in the sense of particle physics. Instead of that, it emerges from quantum field theory as hydrodynamics or continuum elasticity theory emerges from molecular physics.

In this sense, the entropy variation of a system constituted by oscillating quasi-particles (information bits) on a closed holographic screen and that stores the information of matter enclosed within it \citep{navi17} induce gravity. 
As already commented above, DGT expands the Entropic Gravitational Theory \citep{verl11} to low temperatures. In the same sense that Debye extends the Dulong-Petit law to low temperatures.
The low-temperature region is the scenery of the dynamics of galaxies, and they can be described by an equation obtained from first principles by \cite{navi17} and expressed for systems with spherical symmetry as 

\begin{equation}
a\mathcal{D}_1\left(\frac{T_D}{T}\right)= \frac{GM}{R^2},
\label{eq:main3}
\end{equation}
here $T_D$ (Debye temperature) is the only free parameter of the theory (is obtained below), and 
$\mathcal{D}_1\left(\frac{T_D}{T}\right)$ is the Debye first function \citep{deby12}, defined as
\begin{equation}
\mathcal{D}_1\left(\frac{T_D}{T}\right)= \left(\frac{T}{T_D}\right) \int_0^{T_D/T} \frac{x}{e^x-1}dx,
\label{eq:debye}
\end{equation}
that has two limits
\[ \mathcal{D}_1\left(\frac{T_D}{T}\right) =
  \begin{cases}
    1       & \quad \text{if } T_D/T << 1 \text{ (high T)}\\
    \pi^2(T/T_D)/6  & \quad \text{if } T_D/T >> 1 \text{ (low T)}.\\
  \end{cases}
\]

The first case (high T) reproduce the Newton gravitational law, whereas the second limit (low T) can see used to obtain the only free parameter of the theory, the Debye temperature, $T_D$. As in the case of the study of the capacity specific of the solids at low temperatures, the Debye temperature for the gravitation also is obtained from observations. Taking into account the proportionality
 between the temperature and acceleration (Unruh relation), the Debye function at low temperatures and written as

\begin{equation}
\mathcal{D}_1\left(\frac{T_D}{T}\right)=\frac{\pi^2}{6} \left(\frac{T}{T_D}\right)=\frac{a}{a_0}.
\label{eq:DebyeLow}
\end{equation}
 This equation constitute the bond between the Debye temperature, $T_D$, and the acceleration scale parameter, $a_0$, and in the limit of low temperatures
the Eq.~\ref{eq:main3} can be written as

\begin{equation}
a(\frac{a}{a_0})=\frac{GM}{R^2}.
\label{eq:main2}
\end{equation}
This  equation is the root of the MOND theory \citep{milg83a}, it is also known as the deep-MOND regime.

In general (for all range the temperatures) the Debye function can be parametrized by a power function of 
type $(a/a_0)^{\alpha}$,  and  Eq.~\ref{eq:main2} becomes
\begin{equation}
a\left(\frac{a}{a_0}\right)^{\alpha}= \frac{GM}{R^2}.
\label{mainDGT}
\end{equation}
with
\begin{equation}
\alpha = \frac{\log \mathcal{D}_1\left(\frac{a_0}{a}\right)}{\log \frac{a}{a_0}}.
\label{alpha_a}
\end{equation}
The two asymptotically  cases are: 

\[ a\left(\frac{a}{a_0}\right)^{\alpha} =
  \begin{cases}
    a       & \quad \text{if } \alpha = 0 \text{ (high T- Newton )}\\
    a\left(\frac{a}{a_0}\right)  & \quad \text{if } \alpha = 1 \text{ ( low T - deep-MOND)}.\\
  \end{cases}
\]

\begin{figure}
\vspace*{-0.0cm}
\hspace*{0.0cm}
\centering
\includegraphics[width=15.0cm]{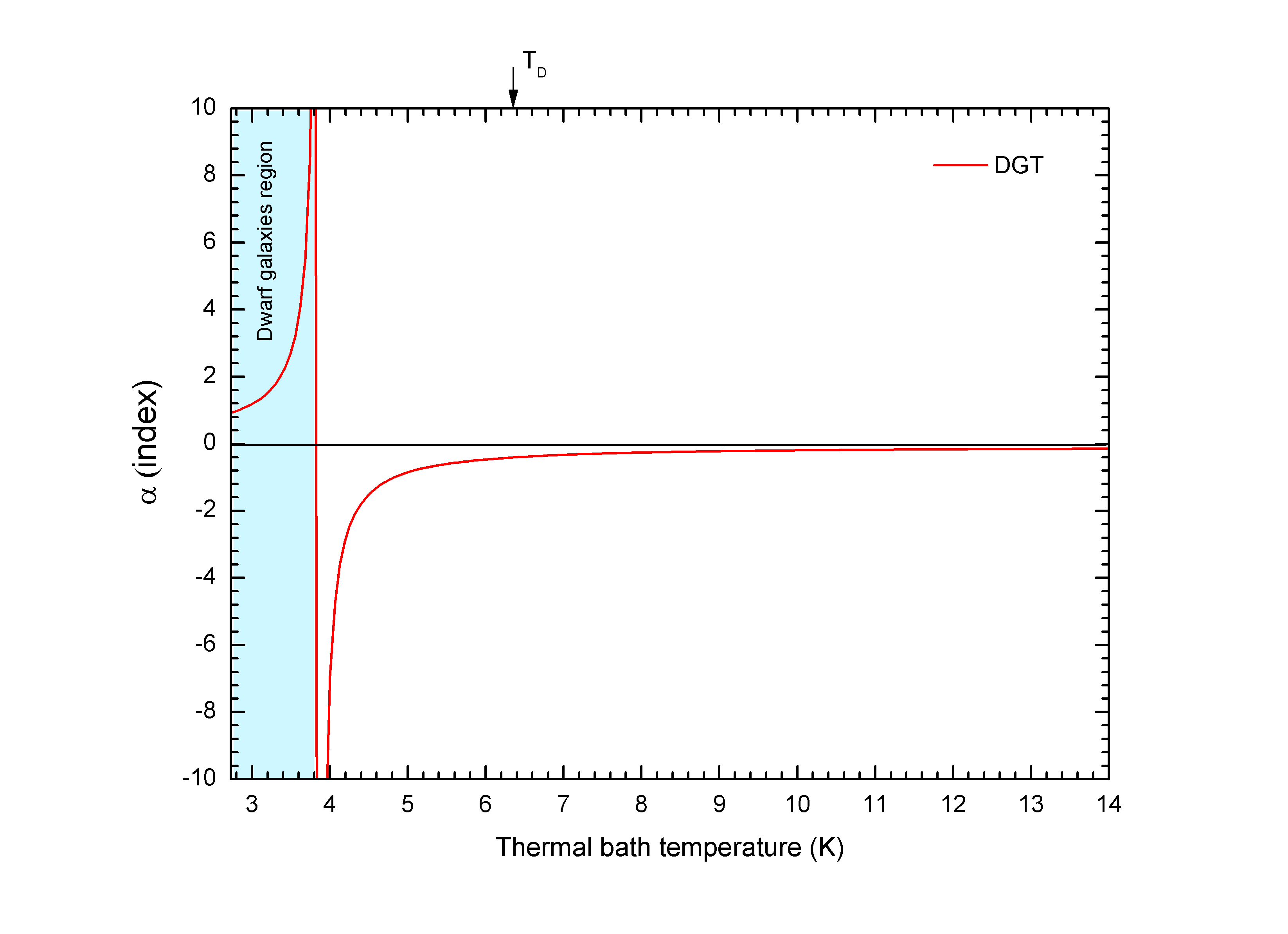}
\vspace*{-0.0cm}
\caption{DGT prediction for the $\alpha$ index as a function of the thermal bath temperature of a galaxy. The vertical arrow at the
upper scale represents the Debye temperature 
($T_D=6.35$ K), and the light blue area is the $\alpha-T$ correlation, supported by the dwarf galaxies.
}
\vspace*{0.5cm}
\label{alpha_dwarf}
\end{figure}

 At present time (redshift z = 0), the thermal bath in which the galaxies are immersed is determined by the temperature of the cosmic background radiation (CMB). Several measurements confirm this temperature as being as $T = 2.73$  K. In DGT this means $\alpha=1$ and the ratio $T/T_D$ for $\alpha=1$ is 
 $(T/T_D)=6/\pi^2(a/a_0)=0.43$, this limit allow us determine the Debye temperature as $T_D=2.73\;K/0.43=6.35$ K.
Fig.~\ref{alpha_dwarf}, shows the dependence of the $\alpha$ index with the temperature.

According to DGT, the light blue area indicates the thermal bath region with temperatures slightly warmer higher than 2.73 K.

We will show that the dwarf galaxies are immersing in thermal baths with these temperatures. We also will show that for temperatures below than the Debye temperature,  the shape
of the rotation curves of the dwarf galaxies,  depend strongly on the temperature of the thermal bath.

\section{DGT predictions for rotation curves of dwarf galaxies}
\label{rotation}

 To make the description of the rotation curves of galaxies, both Newtonian theory and Einstein's general relativity require a hypothetical element, the dark matter, whose nature is unknown, despite the monumental efforts to obtain a direct detection of it, in underground experiments \citep{apri18,agne18} and satellite detectors \citep{albe17}, as well as in the Large Hadron Collider (LHC) at CERN \citep{kahl17}. According to the $\Lambda$CDM the dark matter in a galaxy form a spherical halo that envelops the galactic disc and extends well beyond the edge of the visible galaxy \citep{rubi80}, the halo's mass dominates the total mass.  
 
Another alternative, are the modified theories that mimic dark matter. We highlighted the Modified Newton Dynamics (MOND) proposed by 
\cite{milg83a,milg83b}. MOND is well succeeded, describing the rotation of local galaxies \citep{gent11}, as well as, MOND correlates an accurate measurement of the baryonic Tully-Fisher relation \citep{mcga11}. A review on MOND theory is in \cite{fama12}.
Even so, MOND has some difficulties to explain the rotation curves of dwarf galaxies, such as shown below.

In this paper, we present another alternative. It is a straightforward analysis from DGT, and it needs of two free parameters, the observed galaxy mass and the temperature value to the thermal bath which fit the rotation curve. Thus in DGT, the rotation curve is also an isothermal curve. This analysis already was presented in a previous paper \citep{navi17}. In this paper, we highlighted the rotation curves of dwarf galaxies.

The starting point is the Equation~\ref{mainDGT} that allow obtaining the asymptotic circular velocity as a function of the galaxy radius and taking into account the relation for the centripetal acceleration as $a=\mathrm{v}^2/R$, the rotation speed can be expressed as
\begin{equation}
\mathrm{v}=(GM a_0^{\alpha})^{1/(2\alpha +2)}R^{(\alpha-1)/(2\alpha+2)}.
\label{eq:speed}
\end{equation}

It has two asymptotic limits

%\[ \mathrm{v} =
%  \begin{cases}
%    \sqrt{\frac{GM}{R} }     & \quad \text{if } \alpha=0 \;\;(T/T_D>>1)\\
%    (GMa_0)^{1/4}  & \quad \text{if } \alpha=1 \;\; (T/T_D<<1).\\
%  \end{cases}
%\]

$$\mathrm{v}=\left\{\begin{array}{rc}
\sqrt{\frac{GM}{R} },&\mbox{if}\quad \alpha =0 \;\;(T/T_D >> 1),\\
(GMa_0)^{1/4} , &\mbox{if}\quad \alpha=1 \;\;(T/T_D<<1).
\end{array}\right.
$$
The first asymptotic limit happens at high temperatures, and DGT coincides with the Newtonian regime, and this means Keplerian rotation curves.
According to Fig~\ref{alpha_dwarf}, $\alpha=0$ means galaxies within a thermal bath with a temperature above 13.65 K.
The second asymptotic limit corresponds to low temperatures, and DGT  coincides with the deep-MOND regime, and we have the asymptotically flat rotation curves. According to 
Fig.~\ref{alpha_dwarf}, $\alpha=1$ means galaxies within a thermal bath at a temperature of 2.73 K, this happens for nearby isolate galaxies.

DGT consider that the dwarf galaxies orbiting nearby giant galaxies are within a thermal bath slightly warmer than 2.73 K, this means 
$\alpha > 1$, as is shown in Fig.~\ref{alpha_dwarf}, where the light blue band indicates the correlation $\alpha$-T for dwarf galaxies. 

As already indicated, DGT needs two free parameters, the observed galaxy's mass and the fit parameter, the temperature of the thermal bath in which is immersed ($\alpha$ parameter), to maps the Newtonian rotation curve to the observed one. In this sense, the correlation between the observed circular velocity and the galaxy is

\begin{equation}
\mathrm{v_{cir}}=(GM a_0^{\alpha})^{1/(2\alpha +2)}R^{(\alpha-1)/(2\alpha+2)} -(GMR^{-1})^{1/2}.
\label{circular}
\end{equation}
 
We would like to show, through several examples, that the rotation curves of dwarf galaxies, can be well reproduced by DGT, through 
Eq.~\ref{circular}. We began with the NGC 3109 dwarf galaxy.

\subsection{NGC 3109 dwarf galaxy}

NGC 3109 is a dwarf irregular galaxy, like the small Magellanic nebula. NGC 3109 is the most distant galaxy from the solar system, within the galaxies from the Local Group. NGC 3109 is located about 1.3 to 1.6 Mpc from the solar system, in the Hydra constellation.
NGC 3109 is possibly the fifth largest galaxy in the local group, ranking behind the fourth largest, the Great Magellanic Cloud dwarf galaxy. In addition to suffering the influence of the Milky Way, 
NGC 3109 is not an isolated galaxy, is subjected to tidal effect, 
due to neighboring, the Antly dwarf elliptical galaxy, causing a warp in the disk of NGC 3109. The separation between these two dwarf galaxies is about 40 kpc \cite{greb03}. 

\begin{figure}
\vspace*{-0.0cm}
\hspace*{-1.5cm}
\centering
\includegraphics[width=12.0cm]{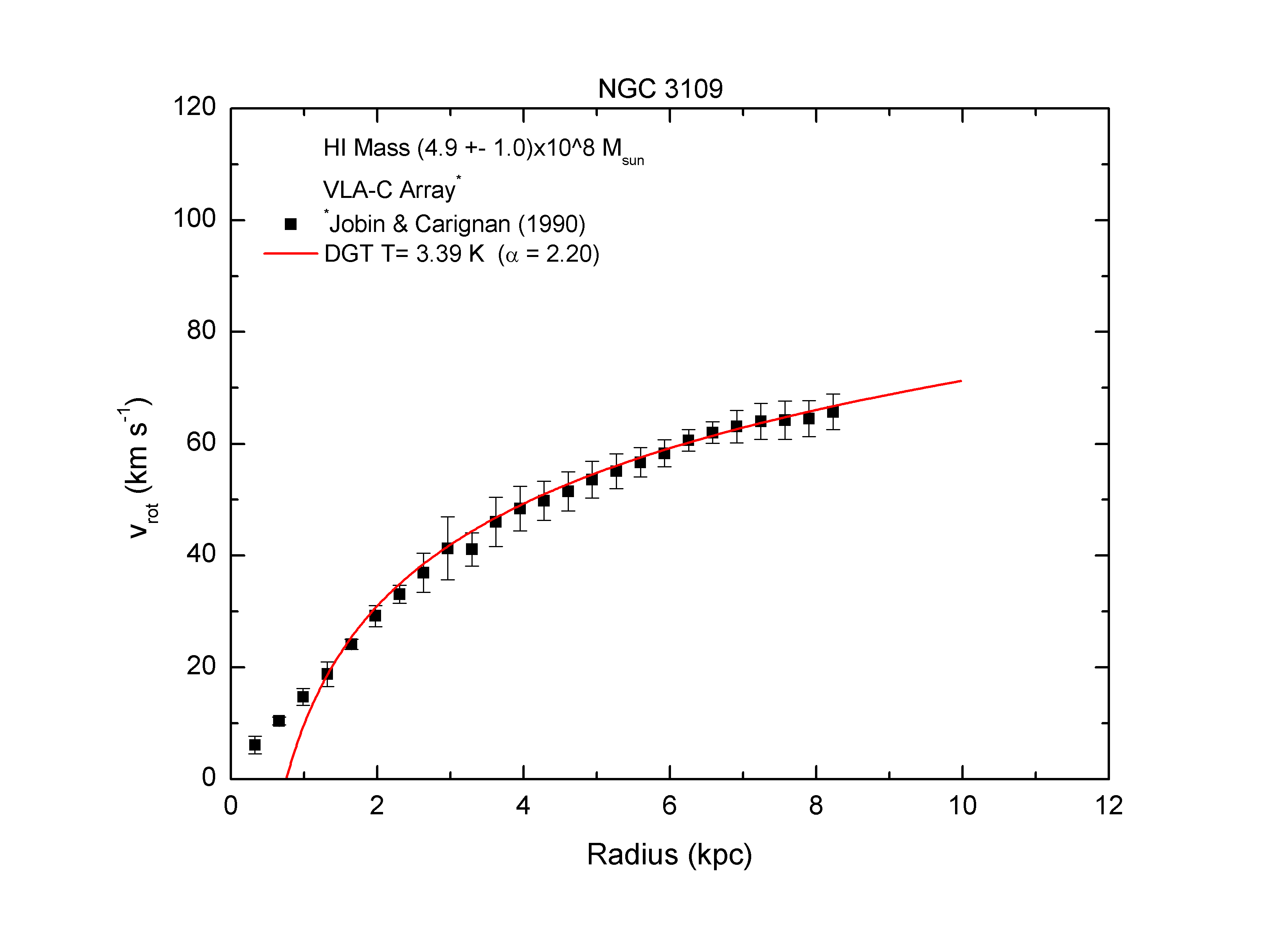}
\vspace*{-0.0cm}
\caption{DGT prediction for the isothermal rotation curve of the dwarf galaxy NGC 3109 (red curve). The squared black points are from VLA observations \citep{jobi90}. The best DGT fit value for the temperature of the thermal bath is 3.39 K.
}
\vspace*{0.0cm}
\label{ngc_3109}
\end{figure}

The solid red curve in Fig~\ref{ngc_3109} shows the DGT prediction for
the rotation curve of NGC 3109 and the black square points are from the VLA observations of \cite{jobi90}.
Except for the three first points at very small radii (r $<$1 kpc), DGT rotation curve is in excellent agreement with the observed curve, and it is like a nonlinear fit on the data, but it is a theoretical prediction from DGT. The thermal bath predicted has a temperature of 3.39 K, i.e.,  0.66 K above the local intergalactic space temperature. 

DGT result is in contrast with the analysis from \cite{jobi90}, from some parametrizations they conclude that NGC 3109 is almost entirely dominated by dark matter. However, the parametrization by a dark halo, require
to reproduce the rotation curve a mass-to-luminosity ratio of $\textit{M}/L_B =0.5$, half to the mass-to-luminosity ratio associated with dwarf galaxies.

There is also an analysis of the rotation curve of NGC 3109 within MOND \citep{bege91}. MOND correlate the data of \cite{jobi90} satisfactorily, with a small deficiency for radios smaller than 3 Kpc. However, also need of a strong requirement for the mass-to-luminosity ratio, it must be quite small, less than 0.1.

\begin{figure}
\vspace*{-0.0cm}
\hspace*{-1.0cm}
\centering
\includegraphics[width=12.0cm]{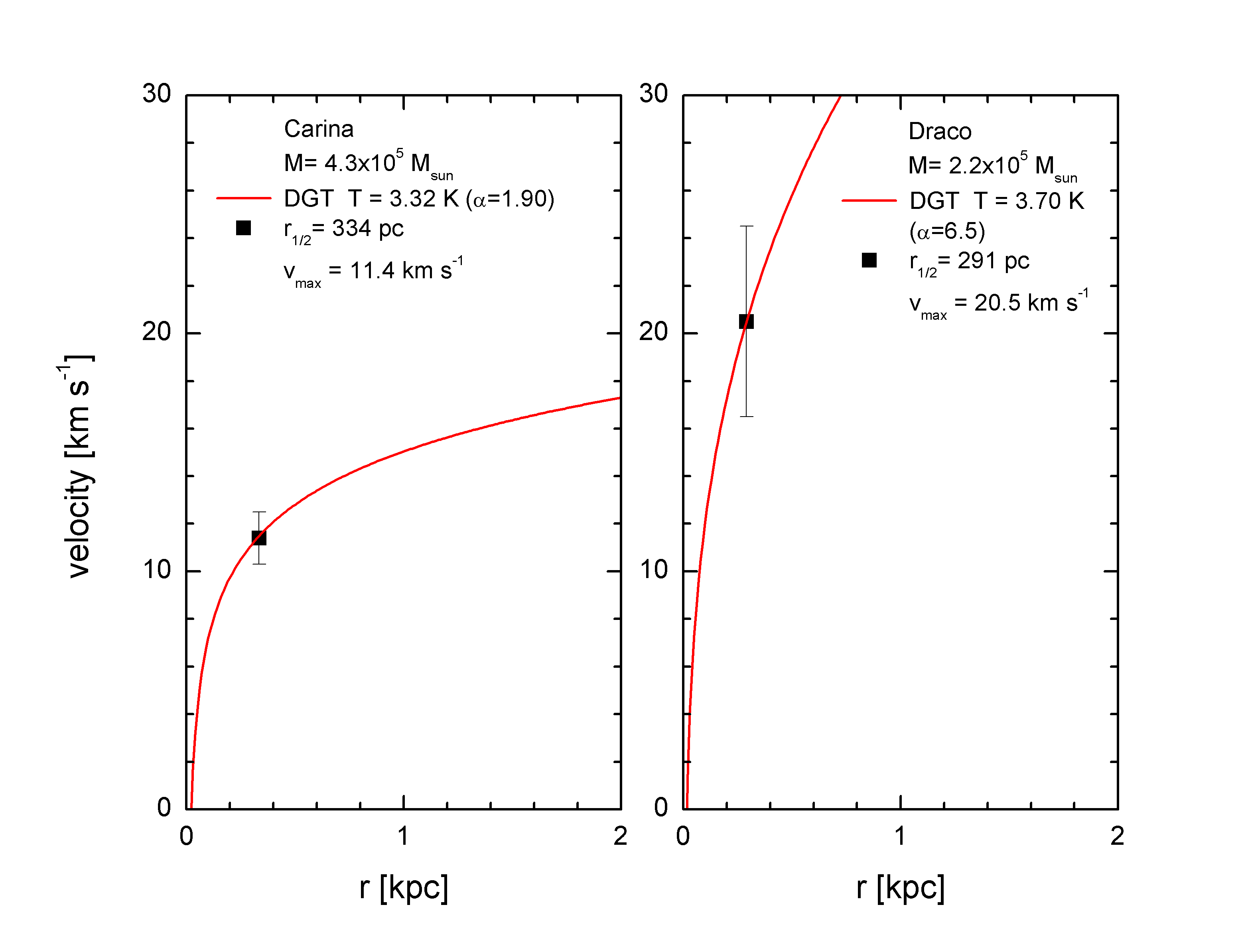}
\vspace*{-0.0cm}
\caption{DGT predictions for the isothermal rotation curves of the dSph galaxies Carina (left panel) and Draco (right panel) (red curves).
The squared black points, represent data from \citep{sawa16}. The best fit values for the temperature of the thermal baths are 3.32 K for Carina and  3.70 for Draco.
}
\vspace*{0.5cm}
\label{draco_rotation}
\end{figure}

\begin{figure}
\vspace*{-0.0cm}
\hspace*{-1.0cm}
\centering
\includegraphics[width=14.0cm]{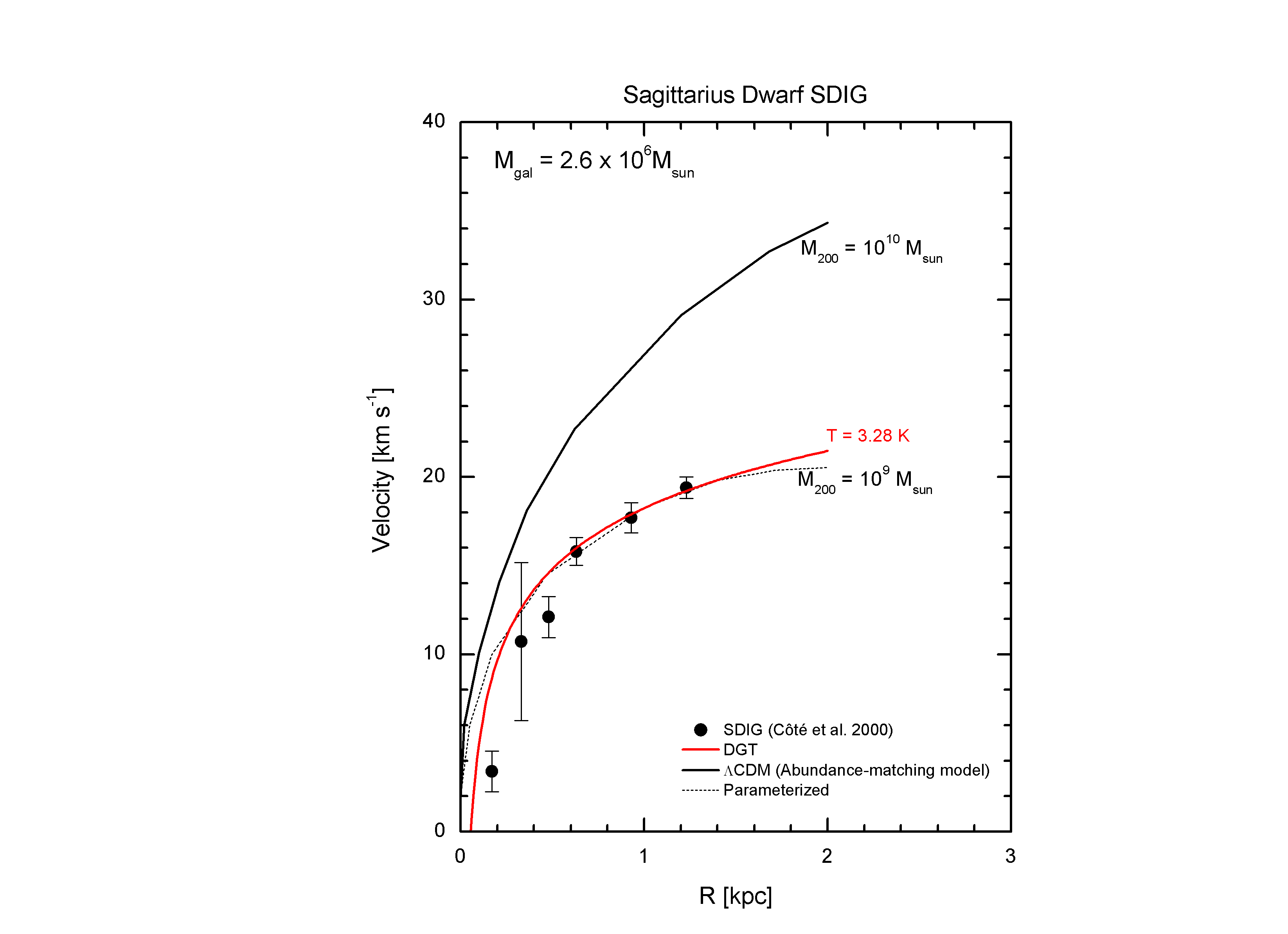}
\vspace*{-0.0cm}
\caption{DGT prediction for the isothermal rotation curve of the dwarf galaxy Sagittarius (SDIG).
The black squared points represent data from \cite{cote00}. The best fit value for the temperature of the thermal bath is 3.28 K. The bold black curve is the prediction from the ``Abundance Matching'' model
\citep{ferr12}.
}
\vspace*{0.5cm}
\label{sagittarius}
\end{figure}

\subsection{Carina and Draco dwarf galaxies}

A challenge for all gravitational models is to reproduce the rotation curves of the faint dwarf spheroidal (dSph) galaxies from Local Group. We have chosen two of them,
Draco and Carina, they are two dSph galaxies, from the Local Group and they are satellite galaxies of the Milky Way galaxy. The Draco and Carina dwarfs are at a distance of 80 Kpc and 100 Kpc from the solar system, respectively. The Carina dwarf has a mass of $4.4 \times 10^5 M_{sun}$, and this value is twice of  Draco's mass $2.2 \times 10^5 M_{sun}$ \citep{sawa16}. Fig.~\ref{draco_rotation} summarizes this information and the  DGT predictions for the rotation curves of these two dSph galaxies. For each galaxy, we have adjusted the temperature of the thermal bath so that the expected rotation curve passes through the point ($ r_{1/2}$ and $v_{max}$).

The data compiled by \cite{sawa16} give a radius of
$r_{1/2}=291$ pc for Draco and $r_{1/2}=334$ pc for Carina, this suggests that Draco and Carina have approximately the same mass profiles as a function of radius. So, they should have approximately the same velocity profiles.  However, according to \cite{boyl12} they
have maximum velocities of rotation very different, $\mathrm{v_{max}}=20.5\;km s^{-1}$ for Draco and $\mathrm{v_{max}}=11.4\;km s^{-1}$ for Carine. This big difference is hard to understand because the observed Draco's mass is half than Carina's mass.
Within DGT, the different velocity profiles of Draco and Carina comes from the difference of temperature of their thermal baths.
 DGT can reproduce Carina's data for a temperature $T=3.32$ K ($\alpha= 1.9$) and $T=3.70$ K ($\alpha=6.50$)  for Draco. Thus, the Draco's thermal bath is 0.45 K warmer than Carina's thermal bath. Figure ~\ref{draco_rotation} summarizes the situation.

Within the $\Lambda$CDM paradigm, this suggested that the amount of dark matter in Draco is more significant than in Carina, result not foreseen by the abundance matching mechanism. 
In this scenario, the galaxy rotation velocities are proportional to halo virial velocity,
and the baryonic mass of a galaxy depends on the efficiency with which halos can condense gas at their centers and form stars. This is known as ``abundance matching'' and it is the principal 
link between the galaxy stellar mass and the halo mass (Guo et al., 2011), expressed as $M_{gal}=f_{bar}M_{200}$, 
where $M_{200}$ indicates the virial halo mass, $M_{gal}$ the galaxy stellar mass and $f_{bar}=0.171$ is the 
universal baryon fraction. However, for $M_{200}$ below $10^{11}$ $M_{sun}$ the 
$M_{gal}-M_{200}$ relation shows a sharp decline in galaxy formation efficiency.
This scenario shows difficulty at small scales, and this is reflected in the rotation curves of dwarf galaxies, especially with masses below than $10^7 M_{sun}$ because in these cases the $\Lambda$CDM is unable to reproduce them. 
This result is a severe constraint the halos of Dwarf Galaxies.

\subsection{Sagittarius dwarf galaxy (SDIG)}

A classic example of the abundance matching mechanism limitation is the prediction of the rotation curve of Sagittarius dwarf irregular galaxy (SDIG) \citep{ferr12}, its rotation curve is shown in Fig.~\ref{sagittarius}. Dwarf Sagittarius has a measured mass is only $2.6\times 10^5 M_{sun}$. The $M_{200}=10^{10}M_{sum}$ condition of the abundance matching mechanism (solid black curve), predict a maximum velocity of rotation about ten times higher than the observed. Fig.~\ref{sagittarius} also shows the
rotation curve (solid red curve) predicted by DGT, and it is seen to agree well with the observed curve. DGT predict a thermal bath temperature for Sagittarius of 3.28 K.

The solution to the abundance matching mechanism of $\Lambda$CDM was the introduction of several 
feedback mechanisms, cold dark matter halos can be heated by supernova explosions and also heated 
through an adiabatic halo contraction \citep{truj11}, or
by busty star formation, increasing the stellar mass-to-halo mass ratio ($M∗/M_{200}$) of dwarf galaxies \citep{read18}.
Besides, the EAGLE an extensive simulation programme based on $\Lambda$CDM \citep{voge13} has been calibrated on small scales to take into account the dwarf galaxies, this new simulation is the so-called APOSTLE \citep{sawa16}. 
The APOSTLE/EAGLE claim that the $\Lambda$CDM simulations both the rotation curves and the TFR relation for dwarf galaxies are in agreement with the data.

%%%%%%%%%%%%%%%%%%%%%%%%%%%%%%%%%%%%%%%%%%%%%%%%%%%%%%%%%%%%%%%%%%%

On the other hand, within the MOND paradigm, the gravitational dynamics of a system is influenced 
by the external gravitational field in which is embedded.
This effect comes from the so-called
External Field Effect (EFE) and is one of the important implications of 
MOND \citep{beke84}. 

According to \cite{fama12} dwarf galaxies, including the faint dwarf satellite galaxies of the Milky Way appear to be in the ``Quasi-Newtonian Regime", defined as $g_{in} < g_{ext} < a_0$. A system in which all accelerations are in the MOND regime but where the external acceleration exceeds the internal acceleration.  In this regime, the effective value of G is enhanced by the factor $a_0/g_{ext}$, but may also show a Keplerian decline in their rotation curve, such as $\mathrm{v}\propto (M/rg_{ext})^{1/2}$ \citep{mcga13}.
 
This Quasi-Newtonian regime of MOND to explain the faint dwarf galaxies of the Milky Way conflicts with the DGT expectation at least for these two dSph galaxies shown in 
Fig.~\ref{draco_rotation}, and also by the dwarf galaxies NGC 3109 and Sagittarius, here analyzed, in all cases, there is a mass discrepancy, but the Keplerian decline in there rotation curves of these galaxies as predicted by the EFE mechanism of MOND is hard to see.

In DGT there are declining rotation curves when the index 
$\alpha$ is negative, in this case,  the thermal bath temperature of the galaxies need a temperature above 4.8 K,
for instance, this happens for galaxies at redshift above 0.77. This DGT prediction \citep{navi18} is in agreement with the VLT observations of declining rotation curves of galaxies with average redshift $z=1.52$ \citep{lang17}. According to DGT, in this case, the galaxies were within a thermal bath of 6.88 K ($\alpha=-0.26$).

 \section{DGT predictions for the Radial Acceleration Relation of dwarf galaxies}
 \label{radial}
 
The radial acceleration relation (RAR) is a correlation between the radial acceleration traced by rotation curves and that predicted by the observed distribution of baryons.
It was obtained for the first time by \cite{mcga16} using the SPARC catalog, a sample of 175 disk galaxies representing all rotationally supported morphological types. In other words, RAR is a scaling-law between the observed acceleration 
$g_{obs}$ and the acceleration generated by the baryonic mass $g_{bar}$ at an any given radius within galaxies.
In this paper, we present only a straightforward analysis from DGT \citep{navi18}, applied to dwarf galaxies.

From measurements of the circular velocity at large radii, we obtain
the observed acceleration as 

\begin{equation}
g_{obs} \sim \frac{\mathrm{v^2}}{R},
\label{RAR_1a}
\end{equation}
while the gravitational acceleration due to the baryonic matter is
\begin{equation}
g_{bar} \sim \frac{GM_{bar}}{R^2},
\label{RAR_2a}
\end{equation}
Taking into account the Eq.~\ref{mainDGT} we can writing it as
\begin{equation}
g_{obs} \left(\frac{g_{obs}}{a_0}\right)^{\alpha} \sim g_{bar};
\label{RAR_3a}
\end{equation}
it can be rewritten  in a more compact form as
\begin{equation}
g_{obs} \sim \left(a_0^{\alpha}g_{bar}\right)^{1/(1+\alpha)}.
\label{RAR_4}
\end{equation}
that has two limits

\[ g_{obs} \sim
  \begin{cases}
    g_{bar}      & \quad \text{if } \alpha = 0 \text{ (Newton)}\\
    \sqrt{a_0 g_{bar}}  & \quad \text{if } \alpha = 1 \text{ (deep-MOND)}.\\
  \end{cases}
\] 

In the log-log scale, the RAR expressed by Eq.~\ref{RAR_4} is a straight line, with slope gives by $1/(1+\alpha)$ and a normalization factor $\alpha/(1+\alpha)\log(a_0)$.
The DGT prediction to the RAR expressed in Eq.~\ref{RAR_4} can be correlated with the data of individual galaxies. For each galaxy is determined the circular velocity at the largest radius (R), to obtain $g_{obs}$
and $g_{bar}$, according to Eq.~\ref{RAR_1a} and Eq.~\ref{RAR_2a}, respectively, where the $M_{bar}$ is the observed galaxy mass. 
This procedure applied to normal local galaxies exhibit
a scaling-law. In DGT this behavior is a consequence of the temperature of the thermal bath be the same 2.73 K ($\alpha=1$) for all local normal galaxies.

However, the thermal bath of dwarf galaxies, are in a range of temperatures from 2.73 K to 3.80 K, where small temperature variations can produce different behaviors. In this cases, an individual analysis can be more useful, and it requires to obtain the 
$g_{obs}(r)$ and $g_{bar}(r)$ at every radius. We believe that this procedure is also more useful when is analyzed a reduced number of galaxies. If the galaxy rotation curve is known, the observed acceleration at every radius is 
\begin{equation}
g_{obs}(r)=\frac{(\mathrm{v_{obs}}(r))^2}{r}.
\end{equation}
However, $g_{bar}(r)$ requires more careful analysis, especially for irregular dwarf galaxies. We have tempted to obtain it, using the method reported in \cite{li18}, but this procedure requires additional information not available to dwarf galaxies. We will show that an effective procedure to irregular dwarf galaxies is to consider an exponential increase of the  galaxy mass with the radius, such as
\begin{equation}
g_{bar}(r)=\frac{GM}{r^2} (1-\exp(-r/r_0));
\end{equation}
the $r_0$ parameter is obtained for each galaxy, through the following condition, at the largest observed galaxy radius, (R), we have the constraint condition 
$(1-\exp(-R/r_0))\sim 1$.

In the RAR analysis, the galaxy mass does not appear explicitly, is embedded in the 
$g_{bar}$ expression. This behavior of RAR can be useful for testing gravitational models that describe a galaxy. Besides the rotation curve, i.e., the circular velocity as a function of galaxy radius, a gravitational model should also to reproduce the RAR as a function of galaxy radius. In DGT the correlation between $g_{obs}$ and $g_{bar}$ at any galaxy radius, is obtained using the circular velocity  expressed in 
Eq.~\ref{circular}; so the $g_{obs}$ is defined as $g_{obs}=(\mathrm{v_{cir}(r)})^2/r$ and the correlation is expressed as

\begin{equation}
g_{obs} =\left(  \left(a_0^{\alpha}g_{bar}\right)^{1/2(1+\alpha)}-g_{bar}^{1/2}\right)^2.
\label{RAR_5}
\end{equation}
and has two limits
\[ g_{obs} \sim
  \begin{cases}
    g_{bar}      & \quad \text{if } g_{bar}>a_0 \text{ (Newton)}\\
    \left(a_0^{\alpha}g_{bar}\right)^{1/(1+\alpha)}  & \quad \text{if } g_{bar}<a_0 \text{ (DGT)Eq.~\ref{RAR_4}}.\\
  \end{cases}
\] 

\begin{figure}
\vspace*{-0.0cm}
\hspace*{0.0cm}
\centering
\includegraphics[width=15.0cm]{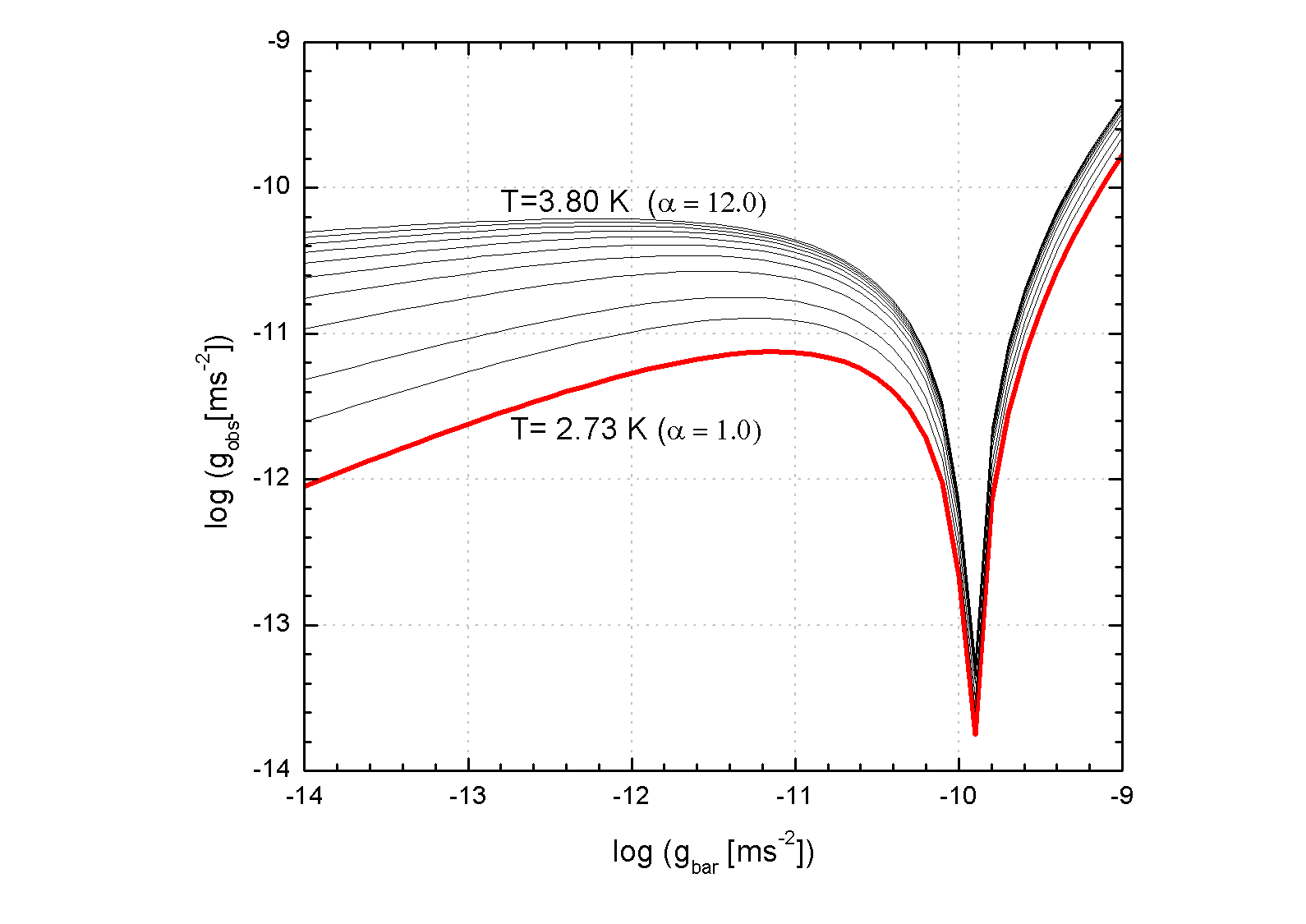}
\vspace*{-0.0cm}
\caption{DGT prediction for the isothermal curves for the RAR of individual dwarf galaxies. The bold red curve represents the deep-MOND prediction ($\alpha=1$ in DGT) and the black curves represent the DGT predictions, from $\alpha=1.5$ to $\alpha=12$, this is equivalent to consider, thermal bath temperatures from T=2.73 K to
T=3.80 K. 
}
\vspace*{0.5cm}
\label{RAR_VLT}
\end{figure}

The last limit, for $\alpha=1$ reproduce the RAR according to deep-MOND prediction ($g_{obs}=\sqrt{a_0g_{bar}}$). In all cases, i.e., for all possible temperature ($\alpha$ values), the $g_{obs}$ expressed by Eq.~\ref{RAR_5}
tends to a minimum value, when $g_{bar} \sim a_0$. This mark the separation of two physical regimes,
the DGT (including deep-MOND) regime at left, and a quasi Newtonian regime at right,
this last, as $g_{bar}$ increases, goes asymptotically to a Newtonian regime where $g_{bar}\rightarrow g_{obs}$.
Fig.~\ref{RAR_VLT} summarizes the situation, where the bold red curve, indicate the RAR for nearby galaxies, immersed in a
thermal bath at 2.73 K. The eleven black curves are for galaxies, immerses in thermal baths with temperatures from 3.2 K to
3.8 K. The rising of the RAR curves with the temperature tend to a saturation. This behavior is a consequence of the $\alpha-T$ correlation (Fig.~\ref{alpha_dwarf}), as the $\alpha$ index increases, the temperature goes asymptotically to a value of 3.8 K.
In all cases, the RAR for dwarf galaxies ($\alpha > 1$) is above the RAR scaling-law (bold red curve). The $g_{obs}$ is systematically higher than $g_{bar}$ and this difference increases as the thermal bath of the dwarf galaxy increases.

%%%%%%%%%%%%%%%%%%%%%%%%%%%%%%%%%%%%%%%%%%%%%%%%%%%%%%%%%%%%%%%%%%%%%%%%%%

\begin{figure}
\vspace*{-0.0cm}
\hspace*{0.0cm}
\centering
\includegraphics[width=17.0cm]{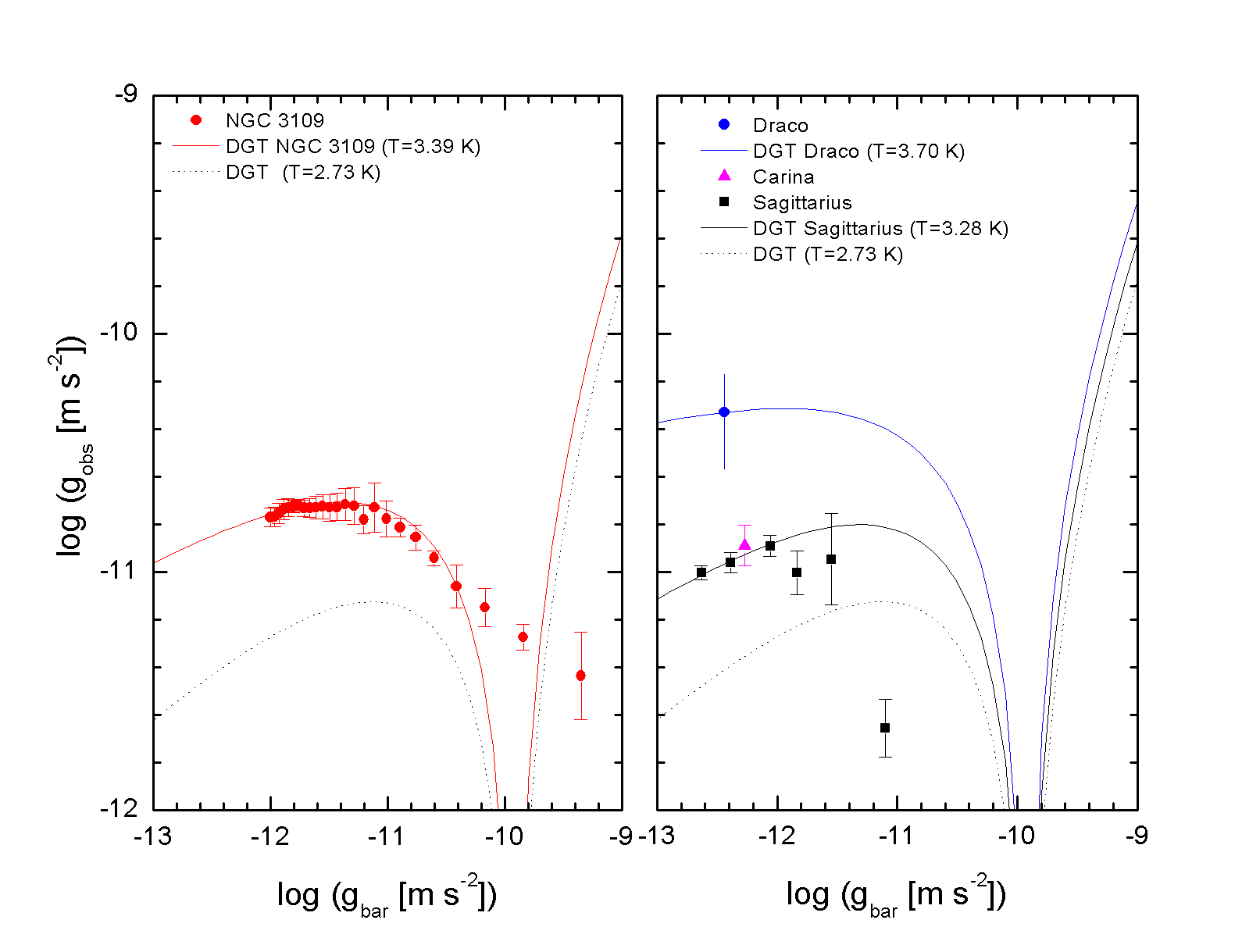}
\vspace*{-0.0cm}
\caption{Comparison of DGT predictions to the isothermal curves of the RAR of individual dwarf galaxies and the data 
of the dwarf galaxy NGC 3109 (left panel) and the data of the dwarf galaxies, Sagittarius,
Carina and Draco (right panel).
}
\vspace*{0.5cm}
\label{RAR}
\end{figure}

We also compare the RAR (individual) for the two dwarf irregular galaxies NGC 3109 and Sagittarius (SDIG) and the predictions from DGT (Eq.~\ref{RAR_5}).
Fig.~\ref{RAR}, left and right panels show the results.

This analysis we extend for the two dSph galaxies Carina and Draco. However, now we have only a  point ($g_{bar}, g_{obs}$) for each galaxy, those for the maximum circular velocities  (see Fig.~\ref{draco_rotation}), the results are compared with the RAR predicted by DGT (Eq.~\ref{RAR_5}). The results are shown in Fig.~\ref{RAR}(right panel).

It should be noted that the  two galaxies, the dwarf irregular Sagittarius (SDIG) and the dSph Carina,
have very different structures, different measured masses, and sizes, but they are correlated in DGT by the same RAR function, expressed by Eq.~\ref{RAR_5} that depending only on the $\alpha$ index, that is, of the temperature of the thermal bath.
Indeed, these two galaxies are within thermal baths almost with the same temperature, $T \sim 3.28 $ K for Sagittarius and 
$T \sim 3.32 $ K for Carina, obtained from their rotation curves. 
In all cases, the same temperature that correlates the RAR of dwarf galaxies is also the one, that best correlates their rotation curves.

From these figures is seen that the data for the RAR of dwarf galaxies is well correlated with the RAR expected from DGT. The exception is the three points, those with the smallest radius in the rotation curve of the NGC 3109 galaxy, they are out of the rotation curve predicted by DGT.  Similar situation is observed for the case of the Sagittarius galaxy, however, now there is only a point, the one with the smallest radius, out the rotation curve predicted by DGT.

In general, the good agreement between the DGT predictions, and the observations for the RAR of the dwarf galaxies, the irregular
NGC 3109 and Sagittarius and the two dSph galaxies Carina and Draco is an indication that the dwarf galaxies break the RAR scaling-law observed in nearby normal galaxies.

However, for thermal bath temperatures above 4.9 K (this happen, for instance, for galaxies at a redshift above 0.77), DGT predict an opposite situation,  the $\alpha$ index is negative, the galaxies have declining rotation curves, where $g_{obs}$ is systematically lower than $g_{bar}$.
 
This prediction of DGT is in agreement with the VLT observations \citep{lang17} as shown in \cite{navi18}, where is presented a comparison between the expected RAR according to DGT for galaxies at redshift 1.52 ($T=6.88$ K) and the RAR obtained from VLT observations of the rotation of galaxies at an average redshift of 1.52. 

This VLT result, contrast with those obtained with the
introduction of the so-called, stellar feedback processes in $\Lambda$CDM galaxy-formation simulations.
These calculations allow reproducing the RAR scaling-law in nearby galaxies \citep{kell17} but predict a break in the scaling-law for distant galaxies, but in the opposite direction, for example, at z $ \ sim $ 2, $g_{obs}$ is greater than 
$g_{bar}$.

The VLT observation also constrains some models, such as the Emergent Gravity \citep{hoss18} and Quantised Inertia \citep{mccu17}. Both reproduce the RAR for local galaxies, but, at high redshifts, $z\sim 2$, they predict a $g_{obs}$ higher than $g_{bar}$.

%%%%%%%%%%%%%%%%%%%%%%%%%%%%%%%%%%%%%%%%%%%%%%%%

\section{DGT predictions for the steep and faint end of the TFR}
\label{TFR}

On 1977 \cite{tull77} found a scaling relation between the luminosity L and the maximum rotation velocity $V_{max}$ of spiral galaxies as 
$L \propto V_{max}^a$, where $a$ is a number close to 4. The rotation velocity of spiral galaxies is usually measured using the (HI gas) 21 cm line of hydrogen, that can extend much further than the stellar disk \citep{bell01}, the luminosity correlated with the stellar mass, so the TFR also correlated the stellar mass with their rotation velocities
The TFR is useful to measure cosmological distances
in the Gpc region. However, TFR is useful to constrain galaxy formation and their evolution scenarios \citep{vogt96,vogt97,cour97}.

In this paper we highlight the TFR to dwarf galaxies, they are responsible by the steep velocity-mass dependence at the faint end of the Tully Fisher relation. The TFR in this sector already has constrained several models, including the most accepted such as the 
$\Lambda$CDM models. The baryonic Tully-Fisher relation
is not well understood in this context because they predict a relation $M \propto V_{max}^3$ \citep{papo17}.
In this scenario, the galaxy rotation velocities are proportional to halo virial velocity and the steep
velocity-mass dependence results from the decline in galaxy formation efficiency with decreasing halo mass needed to
reconcile the CDM halo mass function with the galaxy luminosity function \citep{sale17}, as well as, due to an adiabatic halo 
contraction\citep{truj11}.

 Besides, in MOND scenario, the deviations of not isolated dwarf galaxies from the Tully-Fisher Relation scaling-law is a consequence of tidal effects. MOND requires an additional mechanism, the EFE, i.e., the gravitational effects from neighboring systems, in this case, the giant host galaxy
\citep{mcga13}. EFE would be responsible for putting the dwarf galaxies out of equilibrium, broken the TFR scaling-law observed in nearby spiral galaxies.

From numerical simulations of dwarf satellite galaxies orbiting around giant hosts,  
deviations from the TFR scaling-law, due to the effect of the external field were quantified by 
\cite{brad00} through the ``$\gamma$'' parameter, $\gamma$ is the number of 
orbits a star should make within a dwarf for every orbit the dwarf makes about its host,  
the discrepancy from the bTFR became significant to $\gamma < 8$ and suggested which 
a dwarf might be out of equilibrium, and this happens when the dwarfs tend to become 
non-spherical \citep{mcga10b}.

DGT postulates that rotation of galaxies depends on the thermal bath temperature in the which they are. This scenario is also valid to the dwarf galaxies, in the local Universe it is found that they are orbiting giant galaxies, this means that their thermal bath temperatures are slightly warmer than 2.73 K because they are subject to additional radiation from their host.
  
  Under the assumption of spherical symmetry and following the Eq.~\ref{mainDGT}, allow us calculate the mass of Galaxy as
\begin{equation}
M(R)= \frac{R^2}{G} a (\frac{a}{a_0})^{\alpha}.
\label{eq:massa_c}
\end{equation}
Considering that the acceleration is related as $a=\mathrm{v^2}/R$, we have the final relation
\begin{equation}
M(R)= \frac{R^{1-\alpha}}{G a_0^{\alpha}} \mathrm{v^{2\alpha+2}}.
\label{TFR_DGT1}
\end{equation}
For $\alpha=1$ we have the deep-MOND regime prediction to the mass velocity relation, 
expressed as 
\begin{equation}
M = \frac{1}{G a_0} \mathrm{v^4}.
\end{equation}
This is the well known Tully-Fisher relation, observed in galaxies at very low redshift
(z$\sim$ 0), the slope is in agreement
with the data, while the normalization factor requires a value to the Milgrom acceleration parameter of $a_0=1.2\times 10^{-10}$ m$s^{-2}$ \citep{mcga11}.

In some system such as the elliptic galaxies and dwarf galaxies, only the measurements of the dispersion velocities is possible, and the speed $v$ must be replaced by the velocity dispersion $\sigma$ and the above expression we can write as

\begin{equation}
M = \frac{1}{G a_0} \sigma^4.
\end{equation}
The correlation $M\propto \sigma^4$ is known as the Faber-Jackson relation \citep{fabe76}.

 Returning to Eq.~\ref{TFR_DGT1}, we can see that galaxy mass depends on the radius of the galaxy, the only exception is for $\alpha=1$ (deep-Mondian regime). In general, and for dwarf galaxies, the data is compatible with a power-law dependence, between R and M, parametrized (in the log-log scale) as a straight line 
\begin{equation}
\log{R}=b\log{M}+c,
\label{RvsM}
\end{equation} 
with $b=0.41$ and $c=-2.83$, when R is measured in kpc and M in solar mass \citep{sale17}.
Replacing Eq.~\ref{RvsM} in Eq,~\ref{TFR_DGT1} we obtain the isothermal lines in the TFR diagram
\begin{equation}
\log M=D+F\times \log \mathrm{v}.
\end{equation}   
where the normalization factor D, is given as
\begin{equation}
D=\frac{(1-\alpha)(3+c)-\log(Ga_0^{\alpha})}{1-(1-\alpha)b},
\label{norma}
\end{equation}
the factor (3+c) instead of c in the right side of the equation, transform R from Kpc to pc,
in this case is convenient to use the constants
$a_0$ and $G$  as
$a_0=3.7\;km^2s^-2\;pc^{-1}$ and $G=4.3\times 10^{-3}pc\;M_{sun}^{-1}(km\;s^{-1})^2$,
and the slope F, of the isothermal line of the TFR, is given by

\begin{equation}
F=\frac{2\alpha+2}{1-(1-\alpha)b}.
\label{slope}
\end{equation}
For $\alpha=1$, the normalization factor goes to $D=-\log(Ga_0)$ and the slope to
$F=2(\alpha+1)=4$, i.e., we have the scaling-law of the TFR.

In most cases, the end faint of the TFR is parametrized by a thick line and represent the median values, for instance, for the ΛCDM simulations \citep{sale17,truj11}.  Within the DGT picture, we believe that maybe this procedure is not appropriate, because the end faint of the TFR is the place of a lot of dwarf galaxies, within thermal baths of different  temperatures
of up to one degree above 2.73 K. Thus, DGT predict the isothermal lines (one for each temperature), with slopes and normalizations give by Eq.~\ref{slope} and Eq.~\ref{norma}, respectively.

\begin{figure}
\vspace*{-0.0cm}
\hspace*{-1.5cm}
\centering
\includegraphics[width=14.0cm]{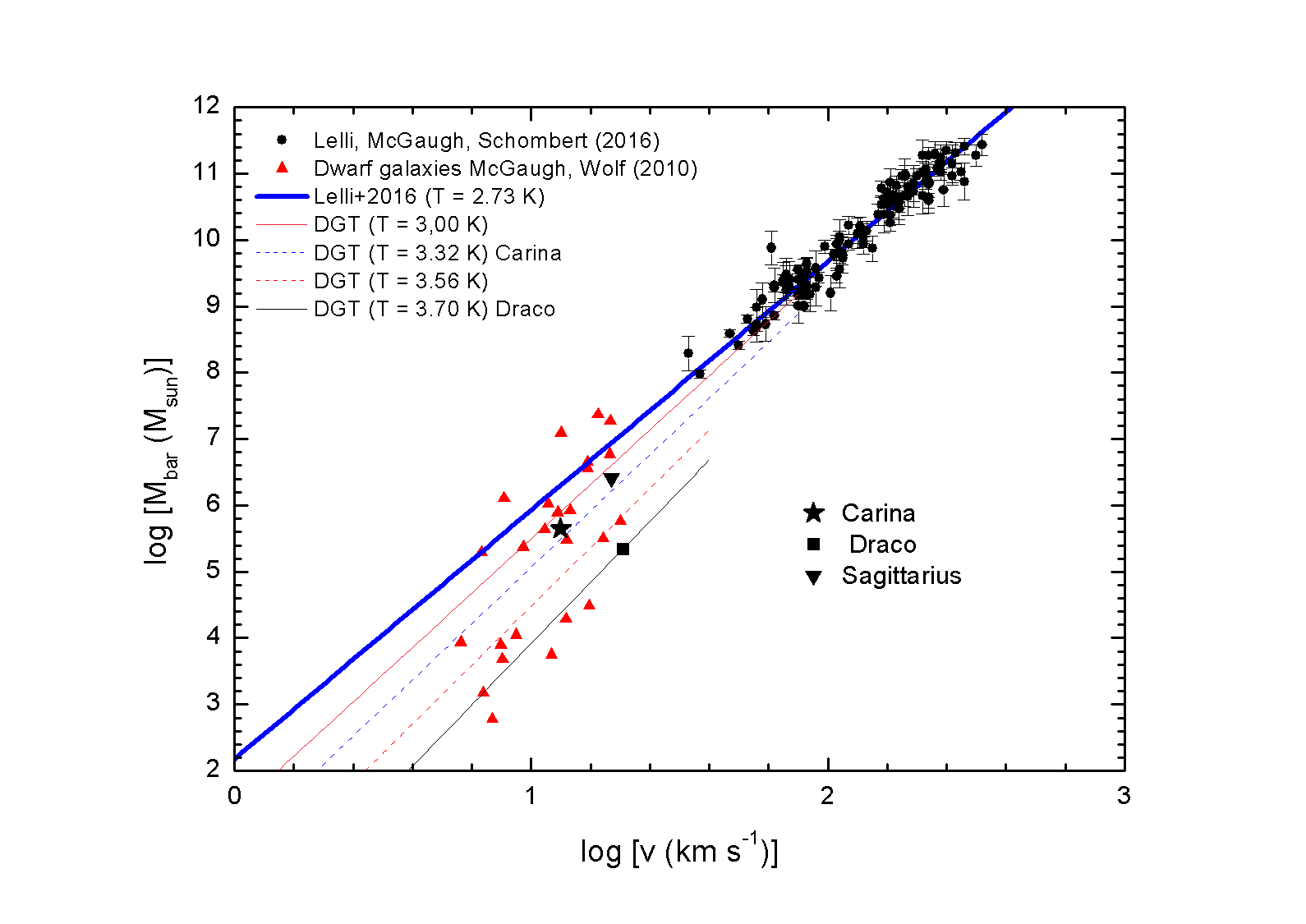}
\vspace*{-0.0cm}
\caption{Comparison of DGT predictions to the isothermal curves of the baryonic TFR of dwarf galaxies and the data from \cite{mcga10} (red triangles). The bold blue line is close with the deep-MOND prediction and it correlates the 
data of nearby no¨rmal galaxies \citep{lell16} (black circles).
}
\label{TFR_B}
%\vspace*{+1.0cm}
\end{figure}

Fig.~\ref{TFR_B} shows the isothermal lines predicted by DGT, including for comparison the data of the dSph galaxies, compiled by \cite{mcga10b}, and the two dwarf elliptical galaxies.

From the rotation curves, are seen that the dSph Carina galaxy and the dwarf irregular Sagittarius galaxy, despite to be very different, they are within thermal baths with temperatures slightly different. This behavior can explain why they can be correlated in the RAR analysis almost by the same isothermal curve (see Fig.~\ref{RAR}, right panel).
This behavior again appears in the TFR analysis, and both fall almost within the same correlation, i.e., the same isothermal line ($T\sim 3.32 K$).

%%%%%%%%%%%%%%%%%%%%%%%%%%%%%%%%%%%%%%%%%%%%%%%%%%

\section{Conclusions}
\label{conclusions}

In this paper, we have presented a new alternative to describe the local dwarf galaxies, from DGT, a thermodynamic gravitational model, inspired in the Debye theory of solids at low temperatures. In this sense, DGT is the extension of the entropic gravitation theory, to low temperatures. In general, we show that DGT has promising results describing dwarf galaxies.
 
The shape of the rotation of any galaxy in DGT depends only on two free parameters the observed galaxy mass and the immersion thermal bath temperature.  DGT uses always the same calibration ($T_D=6.35$ K), presented in section~\ref{background}. This calibration is valid in a wide band of temperatures, for instance, with this calibration DGT predict the rotation curve and RAR of isolated galaxies with an average redshift $z=1.52$, this is equivalent to a thermal bath of 6.88 K. The fall in the rotation curve  predicted by DGT is in agreement with the VLT observations. 
Also with this same calibration,  DGT predicts the dwarf galaxies rotation curves and their radial acceleration relations, including those from the Milky Way. As shown in this paper, they are in good agreement with the observations. 

Nearby spiral galaxies are within a thermal bath at a temperature of 2.73 K ($\alpha=1$), and they have
asymptotically flat rotation curves. For these galaxies, the TFR in DGT is described (in a log-log scale) by a single straight isothermal line with slope, $2(\alpha+2)=4$, and normalization factor, $-\log{Ga_0}$, i.e., a scaling-law. However, dwarf galaxies are immersed in thermal baths with temperatures from 2.73 K to 3.8 K, this means ($\alpha>1$), giving a family
of isothermal lines with slopes higher than 4, breaking the scaling-law. In DGT, this is the origin of the ``knee'' at the end faint of the TFR.  As is shown in this paper, this assumption gives promising results.

However, DGT has its Achilles' heel or perhaps, its major prediction. To $\alpha=-1$
($z\sim 0.77,T=4.83$ K), the rotation curves described by Eq.~\ref{eq:speed} has a divergence. For thermal bath temperatures above 4.83 K, the index $\alpha > -1$ (is negative) and DGT predict decreasing rotation curves, and as the temperature increases, they go asymptotically to a Keplerian one ($\alpha=0$). While to temperatures below 4.83 K, the index $\alpha > 1$ (is positive) and DGT predicts rising rotation curves and as the temperature of thermal bath decreases, they go to a flat curve ($\alpha=1$).

This change in the shape of the rotation curves of galaxies at T$\sim$ 4.88 K predicted by DGT is like as a change in the feature of a physical system, at a given temperature, resulting in a transition of that system to another state, i.e., a phase transition. In other words, when the temperature of the Universe would have reached about $T\sim$ 4.88 K, it would have undergone a phase transition. We want to point out that the temperature of this possible phase transition predicted by DGT is equivalent to a redshift of z$\sim$ 0.77. This value is not far from the redshift value at which the universe would have begun to expand at an increasing rate. According to analyses of Type Ia supernovae, the accelerating expansion of the universe started at a redshift around 0.50 to 0.67. This topic on the possible phase transition of the universe requires more careful analysis, at the moment outside the outline of this work and we will discuss in detail on another occasion.

We want to point out, that all DGT predictions here presented, are easy to reproduce, because they are from a straightforward analysis, they are analytical expressions. So this work contrasts, for instance, with the $\Lambda$CDM predictions from sophisticated simulations.

Also, we have begun to analyze from DGT of the galaxy clusters. The challenge is big, but we already have some promising preliminary results, they will discuss in a separate paper.

\acknowledgments

This work is supported by the Conselho Nacional de Desenvolvimento Cient\'{i}fico e Tecnol\'{o}lgico (CNPq) Brazil, grants 152050/2016-7.

%%%%%%%%%%%%%%%%%%%%%%%%%%%%%%%%%%%%%%%%%%%%

%%%%%%%%%%%%%%%%%%%%%%%%%%%%%%%%%%%%%%%%%%%%%%%%%%%%%%%%%%%%%%%%%%%%%%%%%%%%%%%%%%%%%%%%%%%%%%%%%%%%%%%%%%%%%%%%%%%%%%%%%%%%%%%%%%%%%%%%%%%%%%

\end{document}